 \documentclass[preprint2]{aastex6}

\usepackage{color}
\usepackage{amsmath}
\usepackage{natbib}
\usepackage{wrapfig}
\usepackage{gensymb}
\usepackage{fancyref}
\usepackage{tabularx}

\shorttitle{Chandra Snapshots of Fossil Progenitors}
\shortauthors{Johnson et al. (2018)}

\begin{document}


\title{Chandra and HST Snaphots of Fossil System Progenitors}


\author{Lucas E. Johnson\altaffilmark{1}, Jimmy A. Irwin\altaffilmark{1}, Raymond E. White III\altaffilmark{1}, Ka-Wah Wong\altaffilmark{2}, Renato A.  Dupke\altaffilmark{2,3,4}}

\altaffiltext{1}{Department of Physics \& Astronomy, University of Alabama, Box 870324, Tuscaloosa, AL 35487, USA; e-mail: lejohnson4@crimson.ua.edu}
\altaffiltext{2}{Eureka Scientific Inc., 2452 Delmer St. Suite 100, Oakland, CA 94602, USA}
\altaffiltext{3}{Department of Physics \& Astronomy, University of Michigan, 450 Church St., Ann Arbor, MI 48109, USA}
\altaffiltext{4}{Observat\'{o}rio Nacional, Rua Gal. Jos\'{e} Cristino 77, S\~{a}o Crist\'{o}v\~{a}o, CEP20921-400 Rio de Janeiro RJ, Brazil}



\begin{abstract}
The search for the progenitors to today's fossil galaxy systems has been restricted to N-body simulations until recently, where 12 fossil progenitors were identified in the CASSOWARY catalog of strong lensing systems.  All 12 systems lie in the predicted redshift range for finding fossils in mid brightest group galaxy (BGG) assembly, and all show complex merging environments at their centers.  None of these progenitors had archival X-ray data, and many were lacking high resolution optical data making precision photometry extremely difficult.  Here, we present {\it Chandra} and {\it Hubble Space Telescope} ({\it HST}) snapshots of eight of these strong lensing fossil progenitors at varying stages of evolution.  We find that our lensing progenitors exhibit higher than expected X-ray luminosities and temperatures consistent with previously observed non-lensing fossil systems.  More precise galaxy luminosity functions are generated which strengthen past claims that progenitors are the transition phase between non-fossils and fossils.  We also find evidence suggesting that the majority of differences between fossils and non-fossils lie in their BGGs and that fossil systems may themselves be a phase of galaxy system evolution and not a separate class of object.

\end{abstract}


\keywords{fossil group: fossil progenitors, gravitational lensing: strong, clusters }



\section{Introduction}

Fossil galaxy systems are defined as galaxy systems that exhibit a lack of intermediate brightness galaxies, instead possessing one extremely large central galaxy at their centers.  These fossil systems are thought to have formed these oversized brightest group galaxies (BGGs) over time as dynamical friction slowly drags bright member galaxies down to the BGG, eventually merging with it.  Given enough merging, a $2.0$ magnitude gap in the $r$-band between the BGG and the next brightest member within half the virial radius of the system will be formed, resulting in a classical fossil system,  being born (Jones et al.\ 2003).  These fossil systems were distinguished from lone, large ellipticals by requiring them to possess a hot gas halo of $L_{X, bol}\geq10^{42}h^{-1}_{50}\;$erg s$^{-1}$. Traditionally, this was thought to take a very long time to accomplish given typical angular momentum loss rates via dynamical friction, leading many to believe that fossil systems represented the oldest galaxy systems in the universe.  However, this notion is contradicted by the existence of many nearby fossil systems ($z\lesssim0.1$) that show heated gas in their cores which is inconsistent with these systems being undisturbed (Sun et al.\ 2004; Khosroshahi et al.\ 2004, 2006), since an extremely energetic event is required to heat the intragroup medium (IGM).  Such an event would also need to occur in less than the expected cooling time for these systems, necessitating a recent event implying some fossils may not be as old as previously believed.

Results from N-body simulations by von Benda-Beckmann et al.\ (2008) supported the notion that not all fossils are old structures, as they found many instances of fossil systems being both formed and destroyed since $z<0.9$ indicating this may be a phase of group evolution that all groups have a chance of passing into and out of.  A later study by Kanagusuku et al.\ (2016) found in the Millennium simulation that most groups which are classified as fossils at $z=0$ assembled their BGGs between $0.3<z<0.6$.  These progenitors to today's fossils would be expected to exist near this redshift space and have imminent/ongoing major merging between intermediate mass galaxies and the BGG.  Perhaps coincidentally, this is also the optimal distance away from us for strong gravitational lensing to be possible (Trentham 1995).  This led to the discovery of the Cheshire Cat strong lensing fossil progenitor (Irwin et al.\ 2015).

Identifying the Cheshire Cat as a fossil group progenitor also sheds light on how observed non-cool core fossils might form.  Irwin et al.\ (2015) found the merger of two separate groups could give birth to a fossil system once the BGGs merge, and the shock heating of the hot gas would initially produce a non-cool core.  Since gas cooling time scales can be longer than galaxy merging time scales, one could observe a completed fossil system in a non-cool core phase if it formed via this channel.  Finding the Cheshire Cat also demonstrated that the progenitors to today's fossil systems can be found; to that point little concerted effort had gone into locating these outside simulations (Kanagusuku et al.\ 2016).  After the Cheshire Cat's discovery, we attempted to locate more fossil progenitors in the CAmbridge Sloan Survey of Wide ARcs in the SkY (CASSOWARY) catalog of which the Cheshire Cat (CSWA 2) was a member (Johnson et al.\ 2018).  

Seven classical fossil systems and 12 fossil progenitors were discovered at varying stages of formation giving us some insight into what today's fossils looked like during the formation of their BGGs.  Comparing each fossil category's galaxy luminosity functions (fossils, progenitors, and non-fossils) show the expected differences between fossils and non-fossils and show that progenitors lie between each function demonstrating that they are indeed the transition phase between non-fossils and fossils.  Also discovered was a possible bias for systems acting as strong gravitational lenses to have a higher likelihood of being identified as a classical fossil system when compared to a near identical set of non-lensing control groups.  This, combined with the knowledge that centrally concentrated masses act as better gravitational lenses, suggests that some of the CASSOWARY fossils and progenitors might represent the most extreme examples of fossil systems and/or fossil system formation.

To follow up on these results, {\it Chandra} snapshots were obtained of eight of the twelve newly discovered fossil progenitors in the CASSOWARY catalog at varying stages of BGG formation with the goal of seeing how the hot gas component of a group evolves alongside the galaxy component of a progenitor as it approaches fossil status.  To aid in resolving the innermost regions where major merging is abundant, Hubble Space Telescope ({\it HST}) images in the V, R, and I filters were also obtained for four previously unobserved systems to supplement archival data.  In this work, we report X-ray luminosities and global gas temperatures for eight fossil progenitor systems along with high resolution {\it HST} imaging of fossil BGGs in mid-formation, aiming to gain a more comprehensive understanding of the changes a galaxy group undergoes as the fossil BGG forms and the system concludes its transition to fossil status. 

Section 2 outlines our selection criteria, data reduction, and analysis for {\it Chandra} and {\it HST} images, Section 3 contains our findings using {\it Chandra}, Section 4 our findings from {\it HST}, Section 5 highlights individual systems of interest, and Section 6 summarizes our findings.  We adopt the standard  $\Lambda$CDM cosmology with $H_0=70\;$km s$^{-1}$ Mpc$^{-1}$ and $\Omega_M=0.286$ for all of our equations and figures.

\section{Chandra Analysis}
\subsection{\textbf{Target Selection}}
In our previous work (Johnson et al.\ 2018), we identified 12 Jones fossil progenitors in the CASSOWARY catalog at varying stages of BGG formation.  Sloan Digital Sky Survey (SDSS) images allowed us to observe how the galaxy population of a system evolves as it makes the transition to fossil status on average, with intermediate mass galaxies being cannibalized by the BGG.  This leads to average galaxy luminosity functions shifting toward the faint end as the fossil transition continues, supporting findings from N-body simulations by (Khosroshahi et al.\ 2007).  In addition to using optical light to better understand how the stellar component evolves in fossil formation, X-ray observations help us understand how the hot ICM behaves as a system transitions to a fossil which can give more information as to the recent history of the group.  Of our 12 fossil progenitors, only CSWA 2 was previously observed in X-rays (Irwin et al.\ 2015) severely limiting any general conclusions that could be drawn on the morphology of fossil ICMs.  Moreover, a single progenitor only provides data for a single epoch of evolution.  Ideally, one would want observations of many progenitors at varying epochs of fossil formation to form an observational timeline to supplement existing simulated data.  This motivated our progenitor target selection for follow up {\it Chandra} snapshot observations.

From the 12 previously identified fossil progenitors in the CASSOWARY catalog, we selected eight at different stages in fossil formation.  These stages of evolution correspond to the expected galaxy merger time scales seen in Kitzbichler \& White (2008) and range from 4 Gyr to only $\sim$100 Myr until fossil BGG completion. {\it Chandra} snapshot observations were taken of these eight systems to see how the hot gas component of the groups followed the stellar evolution.  Additionally, aside from one progenitor which shows tentative optical evidence of being an ongoing group merger, each group has comparable richness and mass ensuring we are comparing like systems.  Table~\ref{tab:targets} outlines all our targets along with some basic information about each group.

\begin{deluxetable*}{lcCrlcccc}
\tablecaption{{\it{Chandra} Snapshot Target List} \label{tab:targets}}
\tablecolumns{8}
\tablenum{1}
\tablewidth{1pt}
\tablehead{
\colhead{Target Name} &
\colhead{RA} &
\colhead{Dec} & 
\colhead{$z_{spec}^{BGG}$} & 
\colhead{$N_{200}$} &
\colhead{$M_{200}$} &
\colhead{$R_{200}$} &
\colhead{Time Until} & \\
\colhead{} &
\colhead{} &
\colhead{} &
\colhead{} &
\colhead{} &
\colhead{$\times10^{14}\;$M$_\odot$} &
\colhead{(Mpc)} &
\colhead{Fossil Status (Gyr)}
}
\startdata
CSWA 4	&135.3432\degree&18.2423\degree& 0.346 &32&1.5&1.2&	 0.1									\\
CSWA 10	&339.6305\degree&13.3322\degree& 0.413 &30&1.4&1.1&	3.9   									\\
CSWA 11	&120.0544\degree&8.2023\degree& 0.314 &26&1.3&1.1&		0.2									\\		
CSWA 14	&260.9007\degree&34.1995\degree& 0.442 &18&0.9&0.9&	3.6									\\
CSWA 26	&168.2944\degree&23.9443\degree& 0.336 &83&3.4&1.7&	2.1									\\
CSWA 28	&205.8869\degree&41.9176\degree& 0.418 &31&1.5&1.2&	1.6									\\
CSWA 30	&132.8604\degree&35.9705\degree& 0.272 &31&1.5&1.1&	2.0									\\
CSWA 36	&181.8996\degree&52.9165\degree& 0.266 &26&1.3&1.1&	3.0									\\
\enddata
\tablecomments{A list of our targets along with some basic information about the groups.  $N_{200}$ and $M_{200}$ are measures of the number of member galaxies brighter than $0.4L^*$ and mass contained within the virial radius ($R_{200}$) of the group, respectively.  $N_{200}$, $M_{200}$, and fossil transition time scale is taken from our previous work (Johnson et al.\ 2018).}
\label{targets}
\end{deluxetable*}

\subsection{\textbf{{\it Chandra} Observations and Processing}}
{\it Chandra} ACIS-S snapshots of our eight progenitors were taken during Cycle 17.  Since these were previously unobserved, expected X-ray luminosities and temperatures were found using group scaling relations involving the number of red-ridge elliptical members brighter than $0.4L^*$ in the $r$-band, represented by $N_{200}$ (Lopes et al.\ 2009).  Accounting for redshift and Galactic absorption, fluxes were derived and exposure times chosen to yield a minimum of 100 counts for each progenitor.\footnote[1]{CSWA 26 showed a $2\sigma$ detection in the ROSAT All Sky Survey; this coupled with its high $N_{200}$ motivated us to increase the exposure time to yield an expected 1000 counts.}  The data sets processed uniformly using CIAO 4.7 coupled with CALDB 4.6.9 starting with the level 1 event files following the {\it Chandra} data reduction threads.  Bad pixel files were applied from the standard calibration library in CALDB 4.6.9.

X-ray point sources were found via {\texttt{wavedetec}}t and verified by eye for subtraction, since our work is focused on hot gas emission for these systems.  Curiously, even with our progenitors showing signs of recent and ongoing major merging, we find no active AGN of $L_X>10^{41}\;$erg s$^{-1}$ in any of the member galaxies. Background estimates were found by choosing large regions far from group emission with all background AGN subtracted out.  We chose our fitting regions to be equal to one-quarter of each group's virial radius ($0.25R_{200}$), as the short exposure times made group emission past this point indistinguishable from the background. The tool {\texttt{specextract}} was utilized for spectral work within these regions, including extracting spectra and generating RMF and ARF files for each snapshot.  Spectra were fit within {\texttt{XSPECv12.9}} using an {\texttt{apec}} thermal model incorporating Galactic absorption (Dickey \& Lockman 1990) for each source using {\texttt{tbabs}} and $\chi^2$ statistics as well as the solar abundance table from Grevesse \& Sauval (1998).  Energy channels were grouped so that at least 20 counts were in a single bin with bins ranging from 0.5-7.0 keV.  Any counts below 0.5 kev and above 7.0 keV were ignored due to calibration uncertainties and reduced instrument sensitivity which could introduce unwanted noise.

Spectral fitting yielded X-ray luminosities in the 0.5-7.0 keV energy band with results given in the 0.1-2.4 keV band.  Global temperatures were found via fits for each group, however the relatively poor counts for all but two of our snapshots meant that rough radial temperature profiles were only able to be generated for CSWA 26 and 28.  For these, the total counts within $0.25R_{200}$ were divided equally into two and three concentric annuli for CSWA 26 and CSWA 28, respectively, centered on the X-ray centroid.  The emission within these annuli was extracted and fit, with care taken to have a minimum of 400 counts in each region. Little to no constraints were able to be placed on metal abundances, so we chose to fix it at 40\% solar for this work.  Table~\ref{tab:xray} summarizes our findings.  

\begin{deluxetable*}{lcCcclr}
\tablecaption{Chandra Observation Parameters and Results \label{tab:xray}}
\tablecolumns{6}
\tablenum{2}
\tablewidth{0pt}
\tablehead{
\colhead{Target} &
\colhead{Exposure} &
\colhead{Count Rate} &
\colhead{Net Counts} &
\colhead{$L_X^{0.1-2.4 keV}$} & 
\colhead{$T_{X}$}
  \\
\colhead{Name} &
\colhead{Time (ksec)} &
\colhead{(cts s$^{-1}$)} &
\colhead{} &
\colhead{$\times\;10^{43}\;$(erg s$^{-1}$)} &
\colhead{(keV)}
}
\startdata
	CSWA 4	& 7	&0.024	& 171	& $5.3^{+1.4}_{-1.8}$		& $2.7^{+0.7}_{-0.5}$		\\
	CSWA 10	& 12	&0.013	& 155	& $3.6^{+3.6}_{-1.2}$		& $4.7^{+2.3}_{-1.3}$		\\
	CSWA 11	& 11	&0.029	& 322	& $3.1^{+0.3}_{-0.8}$		& $3.0^{+0.9}_{-0.6}$		\\
	CSWA 14	& 19	&0.010	& 192	& $3.1^{+0.9}_{-0.5}$		& $3.4^{+1.5}_{-0.9}$		\\
	CSWA 26	& 19	&0.043	& 824	& $10.2^{+0.9}_{-0.7}$		& $4.8^{+1.2}_{-0.8}$		\\
	CSWA 28	& 17	&0.068	& 1164	& $17.8^{+0.8}_{-0.8}$		& $7.1^{+1.8}_{-1.3}$		\\
	CSWA 30	& 6	&0.028	& 165	& $3.7^{+0.2}_{-2.3}$		& $3.8^{+1.2}_{-1.2}$		\\
	CSWA 36	& 8	&0.067	& 534	& $6.9^{+0.4}_{-0.4}$		& $4.7^{+1.4}_{-0.8}$		\\
\enddata
\tablecomments{Errors to luminosities are reported at $1\sigma$.  Temperature errors are reported at the 90\% confidence level.  We note that all exposure times (with the exception of CSWA 26) were expected to yield 100 net counts based on the group scaling relations of Lopes et al.\ (2009).  All targets, except for CSWA 26, were 1.5-11.6 times brighter than expected indicating that these systems have more hot gas and have higher gas temperatures than scaling relations predict.}
\label{xray}
\end{deluxetable*}

\subsection{\textbf{{\it HST} Observations and Processing}}
Joint {\it HST} ACS observations of CSWA 11, 14, 26, and 30 were taken to supplement existing archival {\it HST} and {\it Gemini} images of other CASSOWARY catalog members and to help resolve exactly which galaxies go into assembling the eventual fossil BGG.  To match archival data methods and exposure times, three line-dithered exposures were drizzled together using the standard pipeline to form composite images in the F475W, F606W, and F814W bands with total exposure times for each filter coming to around 4100 seconds.  These three bands allow us to create high resolution color composite images of these strong lensing systems and identify complex merging environments that were previously unresolved in SDSS.  To help disentangle overlapping stellar envelopes near the BGGs, we fit each galaxy to both Sers\'ic and exponential profiles using {\texttt{galfit}} which simultaneously finds solutions to multi-component fits for multiple galaxies (Peng et al.\ 2011); our goal is to determine which galaxies are being cannibalized in forming a fossil BGG and thus generate more accurate progenitor luminosity functions for comparisons with previous work by Johnson et al.\ (2018).  


\section{{\it Chandra} Results: X-ray Scaling Relations}

\begin{figure}[h]
\hspace{-0.2truein}
{\includegraphics[scale=0.47]{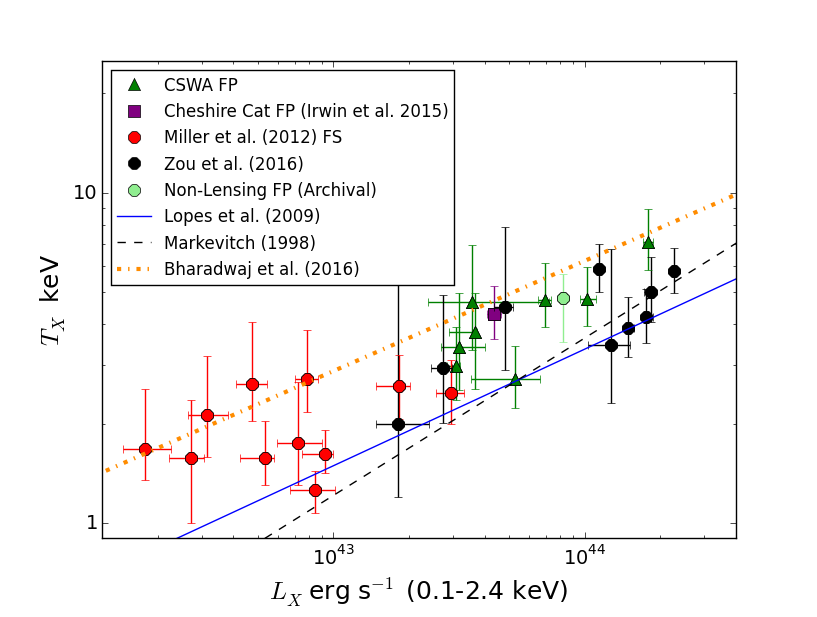}}
\caption{\footnotesize
{The $L_X-T_X$ relation showing strong lensing fossil progenitors (FP) lying near the group scaling relation for fossil systems (FS) found in Bharadwaj et al.\ (2016).  We include the Cheshire Cat system from Irwin et al.\ (2015), as it is also a lensing progenitor in the CASSOWARY catalog.  Non-lensing fossil systems from Miller et al.\ (2012) are also consistent with the Bharadwaj et al.\ (2016) relation.  For better comparison, non-lensing non-fossils from Zou et al.\ (2016) are plotted in black which are consistent with accepted non-fossil group scaling relations.  Archival {\it Chandra} data houses one non-lensing progenitor which is consistent with our lensing sample.  $L_X$ errors are reported at $1\sigma$, and $T_X$ errors are reported at 90\% confidence.}}
\label{fig:L_X_T}
\end{figure}

Lopes et al.\ (2009), using 183 systems, found group scaling relations for radius ($R_{200}$), mass ($M_{200}$), X-ray luminosity ($L_X$), and hot gas temperature ($T_X$) all as functions of the number of $0.4L^*$ red ridge elliptical member galaxies ($N_{200}$).  Earlier studies found similar relations, however these utilized mostly rich clusters (Rykoff et al.\ 2008; Markevitch 1998).  These scaling relations show the expected trend of more X-ray luminous systems housing more bright galaxies along with having higher global hot gas temperatures and are a good way to see if any given subset of galaxy systems deviates from the norm.  For additional comparisons, we include galaxy groups from Zou et al. (2016) who took care to account for any biases in their sample, and we see their systems are consistent with previous group scaling relations formed using more massive clusters, demonstrating the fidelity of our chosen scaling relations.  Our {\it Chandra} snapshots reveal that both nearby, non-lensing fossils (data from Miller et al.\ 2012; Bharadwaj et al.\ 2016),  as well as the CASSOWARY fossil progenitors generally lie above the $L_X-T_X$ relation with our progenitors showing a total temperature elevation of $2.3\sigma$ significance (Figure~\ref{fig:L_X_T}; Table~\ref{tab:xray}).  Bharadwaj et al.\ (2016) found a similar offset using only fossil systems, where their fossils trended above established $L_X-T_X$ relations by a total of $2.3\sigma$.  A search through the {\it Chandra} archive for any data on our non-lensing fossil progenitors identified from our previous study (Johnson et al.\ 2018) yielded one result which has been included in Figure~\ref{fig:L_X_T}.  Surprisingly, this non-lensing fossil progenitor is consistent with our lensing progenitors.

\begin{figure}
\hspace{-0.2truein}
{\includegraphics[scale=0.47]{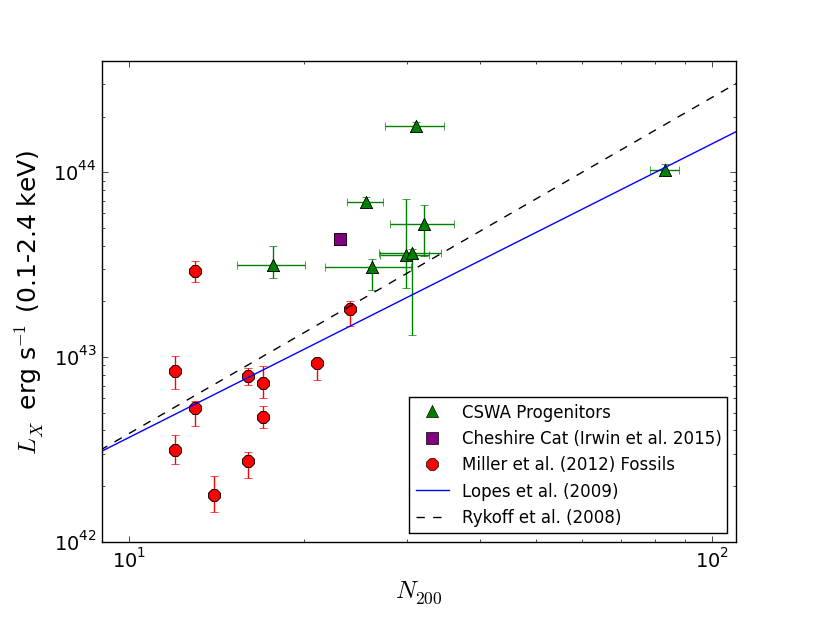}}
\caption{\footnotesize
{The $L_X-N_{200}$ relation showing how strong lensing fossil progenitors are over-luminous by a total of $8.8\sigma$ in X-rays for their bright galaxy count.  For comparison nearby, non-lensing fossil systems from Miller et al.\ (2012) are also included.  Error bars are reported at $1\sigma$.}}
\label{fig:L_X_N_200}
\end{figure}

\begin{figure}
\hspace{-0.2truein}
{\includegraphics[scale=0.47]{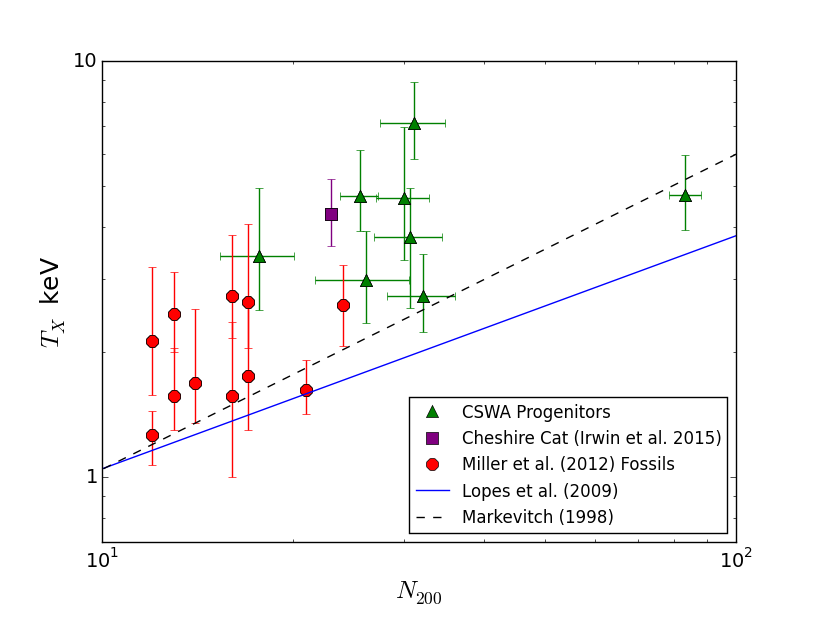}}
\caption{\footnotesize
{The $T_X-N_{200}$ relation also showing an average progenitor temperature elevation of $2.3\sigma$ total significance relative to typical scaling relations.  It is interesting that even non-lensing fossils appear to follow our lensing progenitors suggesting that the lensing bias may not appear in this particular relation.}}
\label{fig:T_X_N_200}
\end{figure}

We also observe a trend among our lensing fossil progenitors of being over-luminous in X-rays compared to their group richness by an average of $8.8\sigma$ significance (Figure~\ref{fig:L_X_N_200}).  This is not seen in the nearby fossils from Miller et al.\ (2012) which could mean this effect is either limited to the progenitor phase, redshift dependent, or is a bias in our sample, as all our progenitors are strong lenses.  One explanation for this behavior is rooted in the definition of a fossil progenitor, namely that they are systems with imminent/ongoing merging.  This merging has the potential to trigger active galactic nuclei (AGN) thereby heating up the surrounding material.  Additionally, we could be seeing the effects of group mergers (e.g. CSWA 2) which would introduce shocks increasing both the temperature and X-ray luminosity of the ICM.  As for the Miller et al.\ (2012) fossils, it was observed that their hot gas halos were often times disturbed and asymmetric which is not expected for relaxed systems.  Therefore, if could be that some observed fossils are young enough to still show the elevated temperatures and luminosities seen in their progenitors.  This possibility is supported by the lack of cool-cores in many nearby fossils (Sun et al.\ 2004, Khosroshahi et al.\ 2004, 2006) indicating that gas cooling time scales can exceed galaxy merger time scales which would produce a non-cool core fossil system for a time.

We see a similar deviation in the $T_X-N_{200}$ relation with our lensing progenitors exceeding typical scaling relations by an average of $2.3\sigma$ significance (Figure~\ref{fig:T_X_N_200}).  However, in this case the Miller et al.\ (2012) fossils seem to follow along with the lensing progenitors in having elevated global hot gas temperatures.  While shocks due to group interactions could cause a system to have elevated temperatures, the likelihood that this is responsible for all fossil system deviations is slim since this effect would be transient.  A lensing bias may be at play here, but if that were true the non-lensing fossils should not follow the same trend as the lensing progenitors.  Instead, it could be that fossil systems and their progenitors possess deeper potential wells than other similar richness non-fossils.  This would serve to both hold on to more group gas and have that gas be at a higher temperature than otherwise predicted (based solely on counting galaxies), however we cannot completely rule out the possibility that our sample is biased.

It is important to note that some deviations in scaling relations using $N_{200}$ can be expected for any given fossil system.  By definition, fossil BGGs are formed via cannibalization of bright member galaxies; this decreases the galaxy count while maintaining the total group gas and stellar mass.  Such deviations would shift fossils left on relations serving to make them appear slightly over-luminous/hotter than predictions. However, this effect alone is insufficient to explain the magnitude of our progenitor offsets, as even the most massive and crowded fossil progenitor in our sample (CSWA 26) will only cannibalize ten $0.4L^*$ galaxies by $z=0$ , representing only a 12\% decrease in $N_{200}$ due to merging.  The other seven progenitors only have between two and five bright galaxies to be cannibalized before fossil status is achieved.  We also find no correlation between time until fossil status is achieved and $L_X$ or $T_X$, however a larger sample size may change this in the future. 

\begin{figure*}
\centering
\hspace{-0.45truein}
\begin{tabular}{lccc}
	{{\includegraphics[scale=0.140]{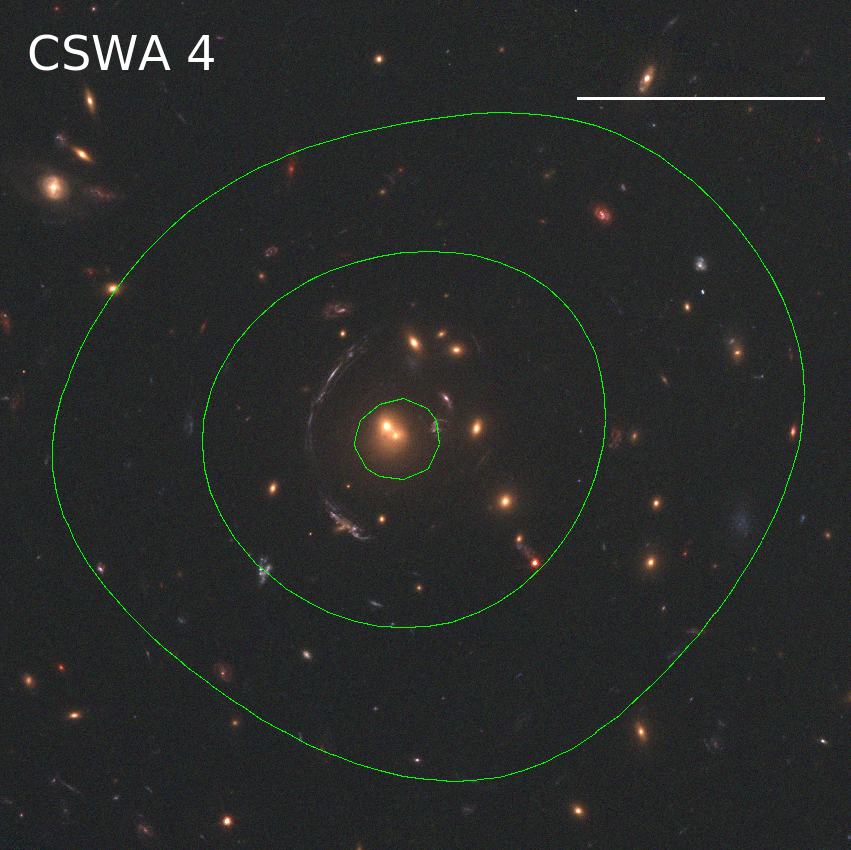}}}&{\includegraphics[scale=0.1303]{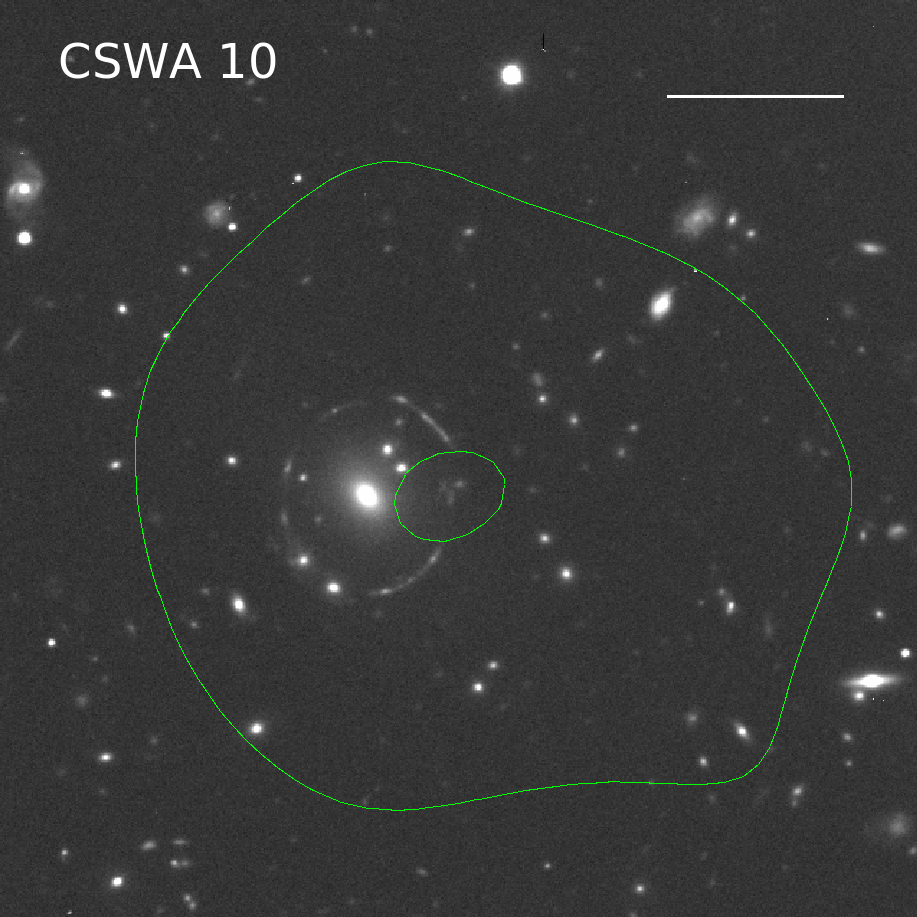}}&{\includegraphics[scale=0.131]{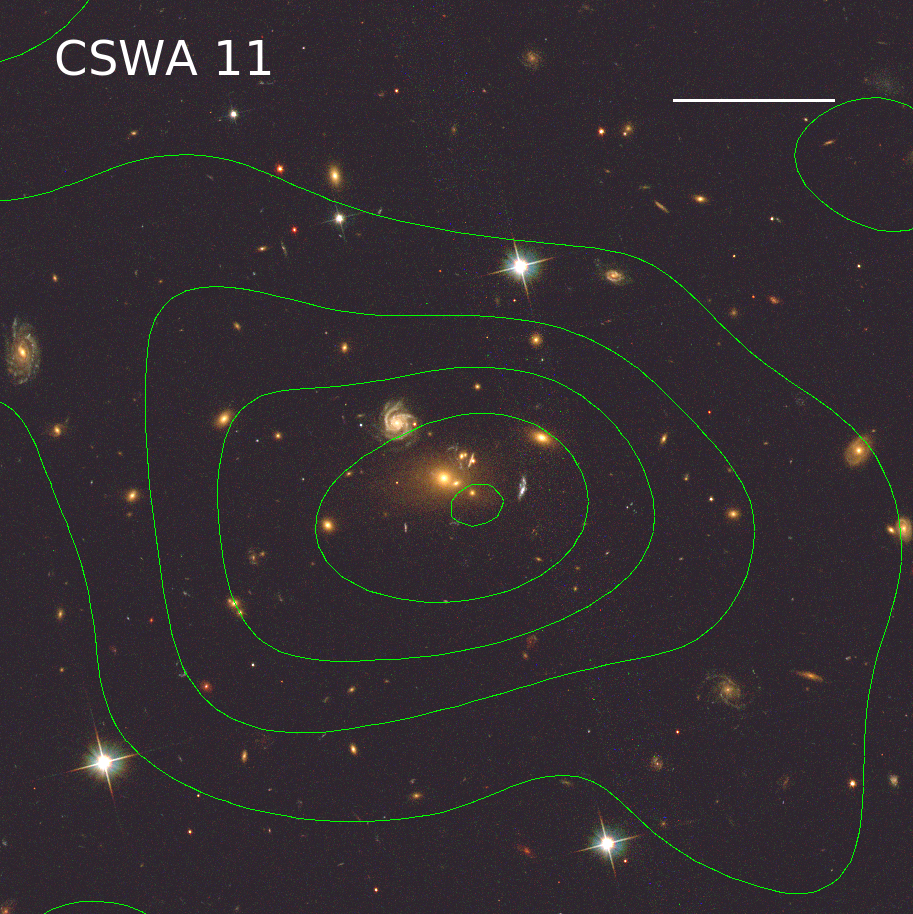}}&{\includegraphics[scale=0.1405]{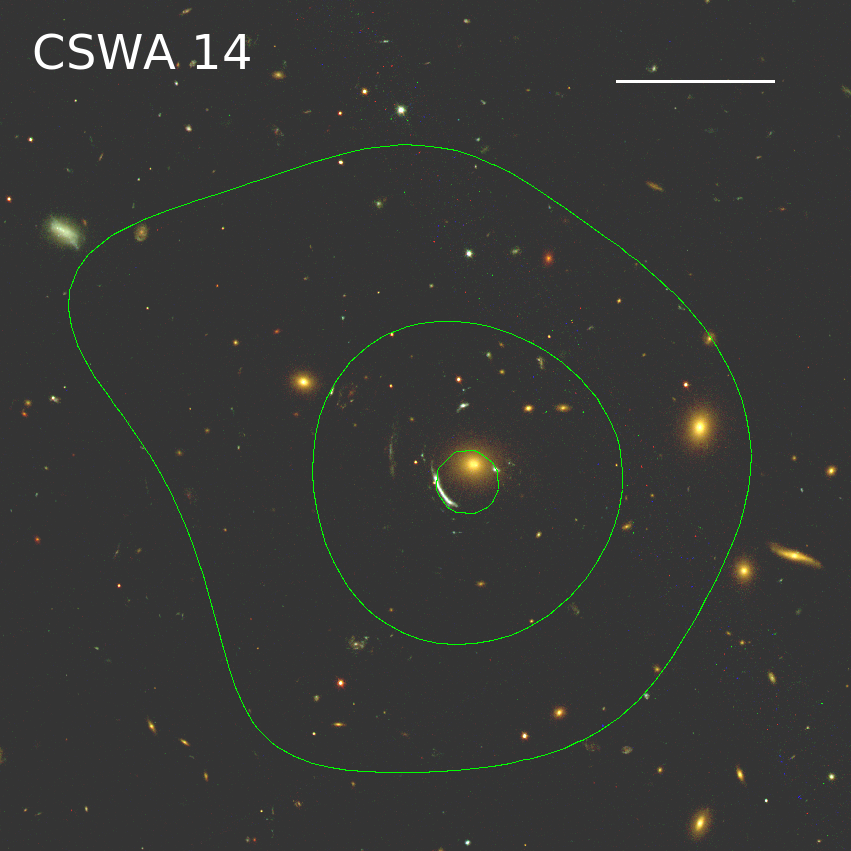}}\\
	{{\includegraphics[scale=0.135]{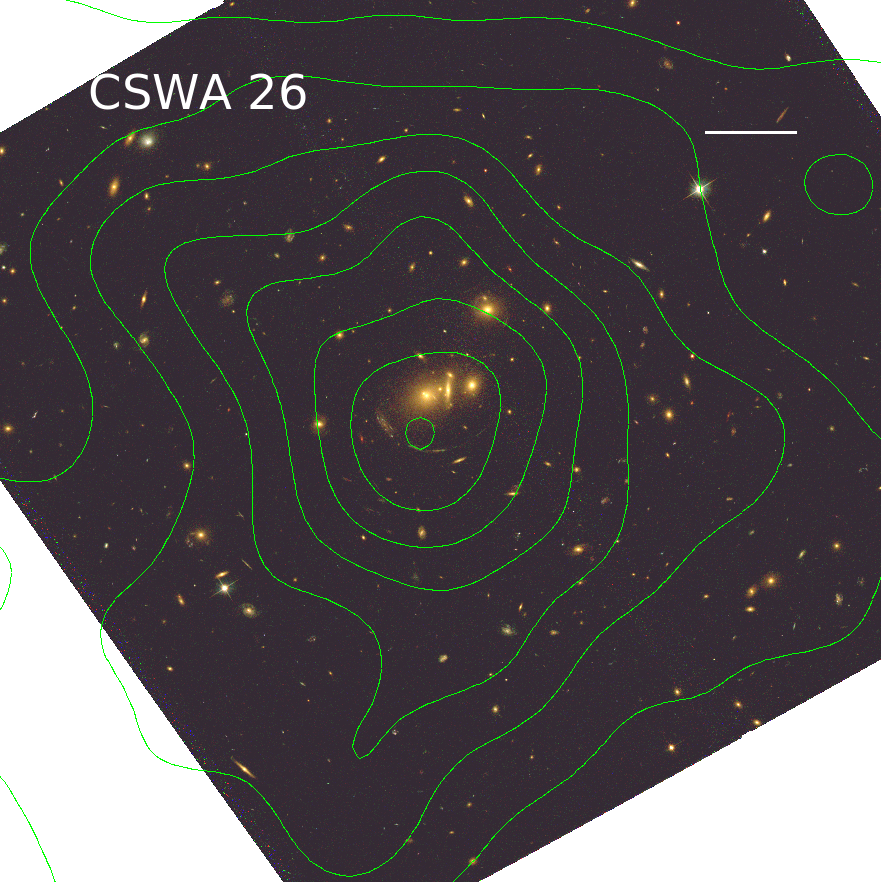}}}&{\includegraphics[scale=0.136]{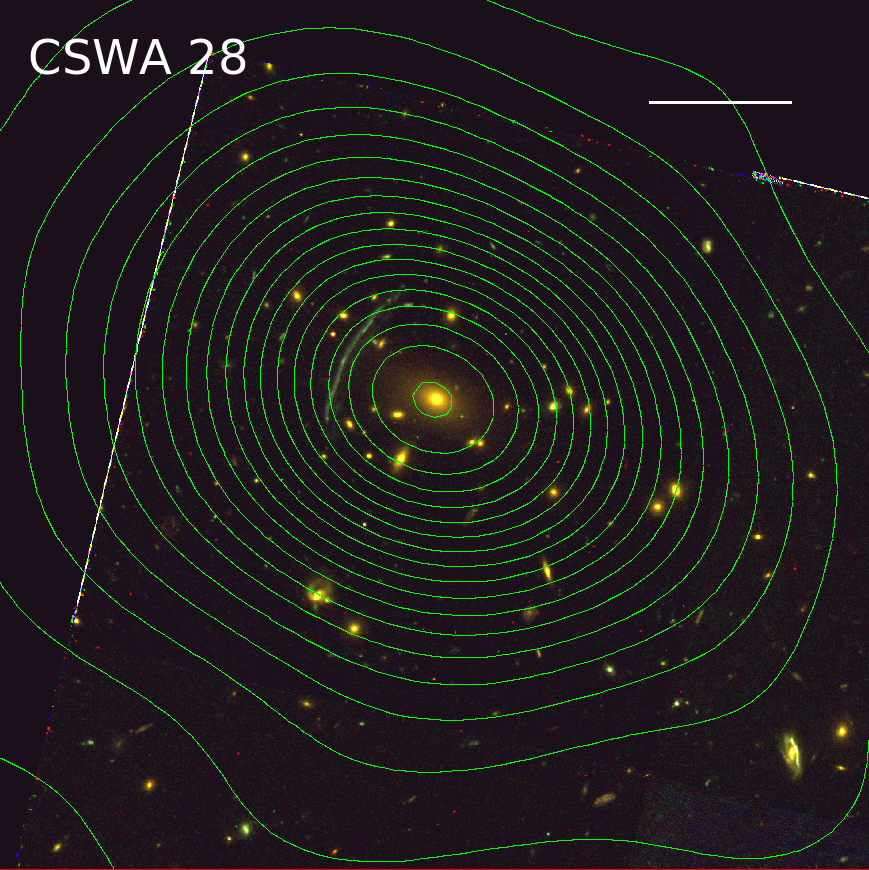}}&{\includegraphics[scale=0.137]{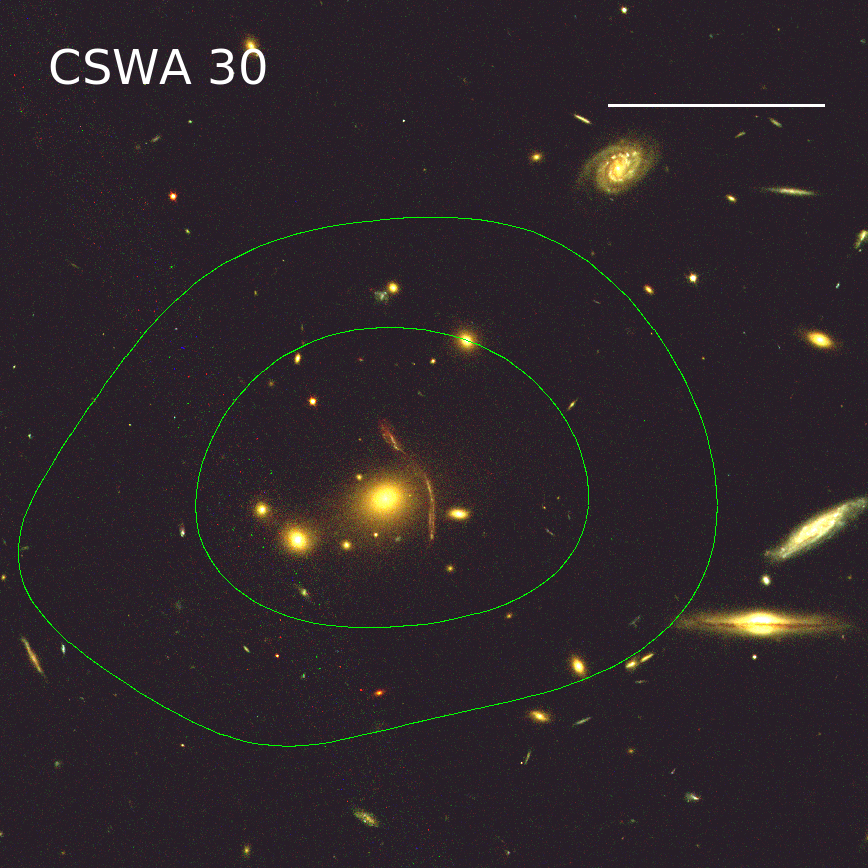}}&{\includegraphics[scale=0.127]{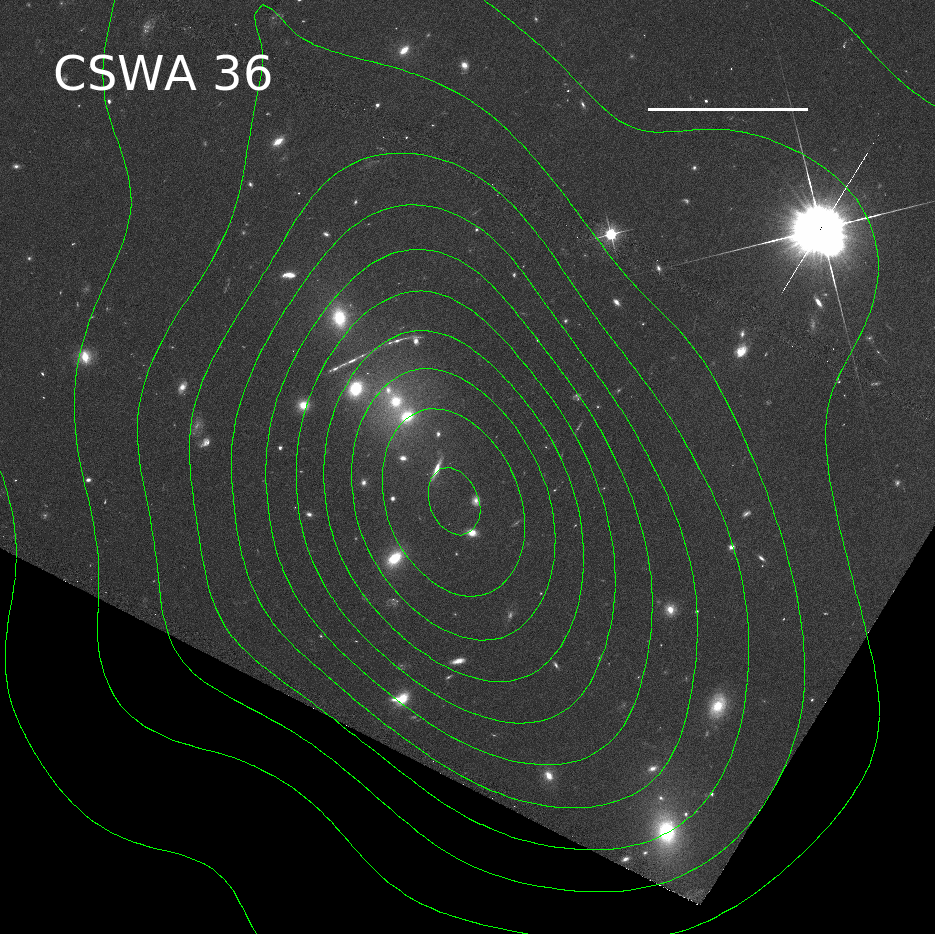}}\\
\end{tabular}
\caption{\footnotesize
{{\it HST} images of the inner regions of seven fossil progenitors (CSWA 10 was observed by Gemini in the $r$-band) with X-ray contours overlaid in green.  Contours begin at $2\sigma$ confidence above the background emission and increase by $1\sigma$.  White bars denoting 100 kpc at each system's redshift are in the upper right corner of each image.  Asymmetric emission is seen in CSWA 11, 26, and 36 with elevated core temperature observed in CSWA 26 and 28.  An offset between CSWA 36's BGG and the X-ray centroid of 60 kpc can also be seen.}}
\label{fig:collage}
\centering
\end{figure*}
\vspace{0.5truein}

\section{{\it Hubble} Results: Improved Progenitor Luminosity Function}

\begin{figure*}
\centering
\hspace{-0.6truein}
\begin{tabular}{cc}
	{{\includegraphics[scale=0.475]{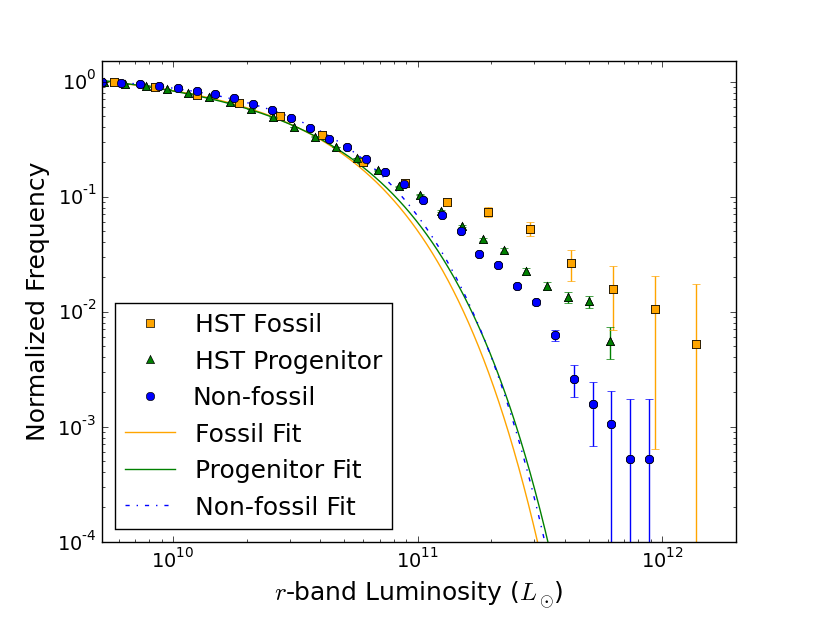}}}&\hspace{-0.5truein}{\includegraphics[scale=0.475]{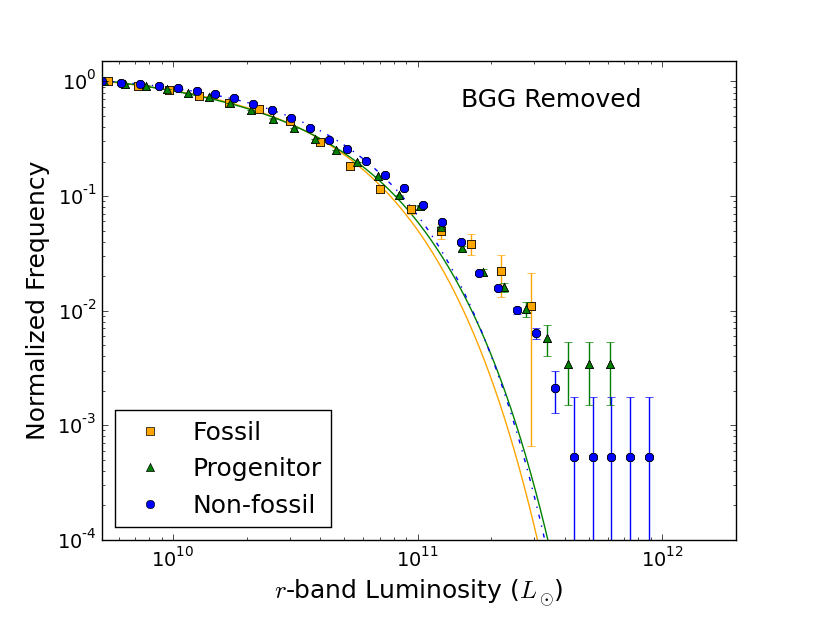}}
\end{tabular}
\caption{\footnotesize
{{\it Left}:  Galaxy luminosity functions of all CASSOWARY catalog members using SDSS photometry refined by all available {\it HST} imaging to resolve BGGs in mid-assembly.  The lines indicate the best fit Schechter functions to the data.  The data diverge near $10^{11}\;L_{\odot}$ clearly demonstrating that progenitors are the transitional phase between non-fossils and fossils.  {\it Right}:  The same data with the exception that all BGGs have been removed.  Here, we see each type of system being consistent with the others; this means that optically the BGG contains most of the differences between fossils and non-fossils on average at this epoch of a fossil's life.  Error bars are reported at $1\sigma$.}}
\label{fig:lum_func}
\centering
\end{figure*}

Previous observations of our CASSOWARY progenitors was limited to SDSS images; and, as our sample's average redshift is $z\sim0.4$, we were unable to resolve exactly what galaxies were in close proximity to the assembling fossil BGG.  This meant our galaxy luminosity functions in Johnson et al.\ (2018) were incomplete at the bright end for the progenitor sample.  Using recently obtained {\it HST} imaging, we now have resolved images of the inner regions of all eight progenitors allowing us to refine the bright end of the progenitor luminosity function.  A Schechter function (Schechter 1976) of the form
\begin{equation}
\Phi = \phi^*(L_{gal}/L^*)^\alpha\;e^{(-L_{gal}/L^*)}
\end{equation}
was used to find the luminosity function where $\phi^*$ is a normalization, $L^*$ is the characteristic galaxy luminosity power-law cutoff, and $\alpha$ is the faint end slope.  We used the {\texttt{galfit}} program to simultaneously fit all interacting/nearby galaxies in the unresolved area for SDSS to Sersi\'c and exponential disk functions in the F606W and F814W bands with the goal being to find the $r$-band luminosity of all galaxies that will be cannibalized by the BGG.  We find that two previously classified fossil systems (CSWA 4 \& 11) using SDSS resolution are in actuality still assembling their BGGs (see Figure~\ref{fig:collage}).  Upon reclassifying these as progenitors and deconvolving all galaxies in the previously unresolved inner regions of our eight fossils, we find the galaxy luminosity function separation between fossil, progenitors, and non-fossils is preserved and even refined compared to our findings in Johnson et al.\ (2018) (Figure~\ref{fig:lum_func}).   Previously, the progenitor luminosity function trended closer to the fossil function at intermediate luminosities.  We believed this was due to us missing intermediate mass galaxies in close proximity to the BGG due to SDSS's angular resolution limit.  By now resolving many new galaxies at the progenitors' centers, the progenitor function now firmly sits between fossils and non-fossils for $L_{gal}>10^{11}\;L_{\odot}$.   Upon removing all BGGs from the luminosity functions, we see all three becoming consistent with one another.  We believe this is a resolution issue in our fossil system sample, since there are less fossil systems than non-fossils leading to higher uncertainties in the luminosity functions.  By combining fossil and progenitors into one data set, the expected differences from non-fossils is seen{\footnote[2]{The best fits for each category are found by excluding BGGs, as these are the result of galaxy reprocessing over time and are consequently not fit well by a Schechter function.}}.  We note that the reclassifying of two previously identified fossil systems as progenitors increased the uncertainties in the fossil luminosity function by a considerable amount, making it harder to distinguish small variances between data sets.  Additionally, we truncate our data at the absolute $r$-band magnitude $M_r<-19.5$ which corresponds to the completion threshold of SDSS at our most distant CASSOWARY group.

\begin{figure}
\hspace{-0.2truein}
{\includegraphics[scale=0.47]{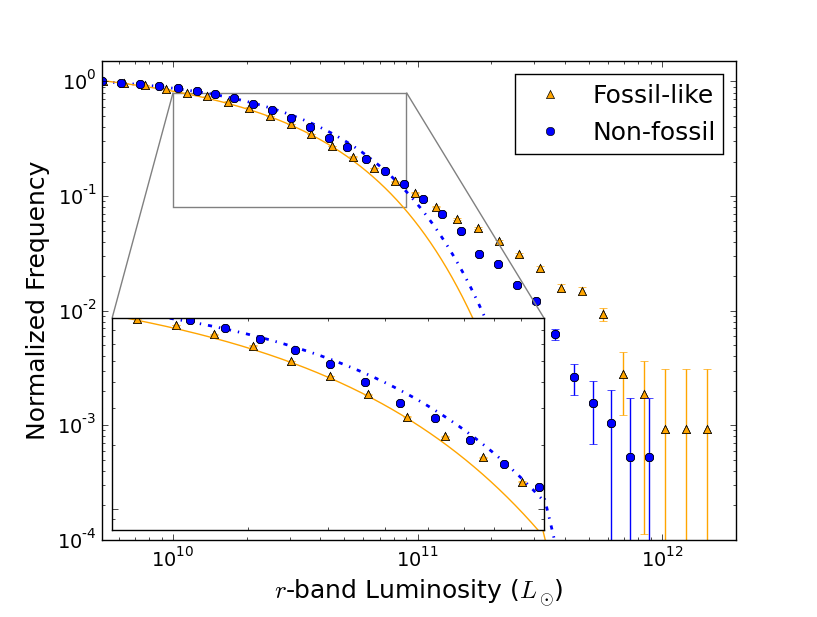}}
\caption{\footnotesize
{The rest frame ($z\sim0.4$) galaxy luminosity functions for a `fossil-like' category (fossils + progenitors) and non-fossils.  While slight, a statistically significant deficit of $10^{10}<L_{gal}<10^{11}\;L_{\odot}$ galaxies can now be seen in the `fossil-like' function which was not previously resolved in this sample (see inset).  This is, in general, expected to grow with time as more galaxies are cannibalized to brighten the BGG.}}
\label{fig:rest}
\end{figure}

In our previous work (Johnson et al.\ 2018), we were unable to resolve any significant $L^*$ deficit in our fossil system luminosity function due to there being a relatively small number of fossils compared to non-fossils.  However, by combining $z\sim0.4$ fossils and progenitors into one `fossil-like' category, a slight but significant deficit is observed between $10^{10}<L_{gal}<10^{11}\;L_{\odot}$ (Figure~\ref{fig:rest}). While less significant than what is expected in nearby ($z<0.2$) fossil systems due to the expected, rapid widening of the BGG magnitude gap at this epoch (Gozaliasl et al.\ 2014), this small deficit should grow as $z$ decreases as more galaxies are cannibalized by the BGG thus widening both the bright end and faint end deviations from non-fossils as fainter galaxies disappear from the population and the bright BGG increases in luminosity.  We also notice the best fit parameter $L^*$ is slightly lower for the `fossil-like' fit compared to the non-fossil fit, however it is not statistically significant.  It could be that this separation in $L^*$ increases with deceasing redshift as more galaxies are cannibalized by the BGG, as Zarattini et al.\ (2015) found that fossils at $z<0.25$ have characteristically lower $L^*$ values than non-fossils.  They go on to find differences for the faint end slopes of the Schechter functions, however the higher average redshift of our groups limits our ability to fully account for faint member galaxies.
 
Since we know which galaxies have the potential to be incorporated into the still forming fossil BGG, we can fast forward each system to a $z=0$ frame (where all our progenitors would be considered fossils) and then compare the new fossils against our projected $z=0$ non-fossils.  By adopting the conservative merger time scale used in Kitzbichler \& White (2008), who utilized the projected separation and total mass of galaxies to estimate the time until two galaxies merge, we identify all member galaxies that could merge with the BGG within the system's look back time add their luminosities to the BGG to create a probable $z=0$ BGG.  After this, we find that the $L^*$ discrepancy between fossils and non-fossils grows slightly ($L^*_{fossil} = 3.3\times10^{10}\;L_{\odot}$ and $L^*_{non}=2.9\times10^{10}\;L_{\odot}$) and is significant, indicating that fossil and non-fossil BGGs must grow at roughly similar rates. This appears to contradict findings in the Millennium simulation by Gozaliasl et al.\ (2014) who found that fossil BGGs rapidly grow their magnitude gap beginning around $z\sim0.2$ implying a simultaneous rapid depletion of other member galaxies.  If this is true, in order for the fossil/non-fossil luminosity function difference to exist, the fossil BGGs must have formed earlier than their non-fossil counterparts.  This supports other findings by Gozaliasl et al.\ (2014) who saw that on average, fossil BGGs assembled most of their mass before $z=0.5$.  One possible cause for the slower than expected growth of fossil BGGs (or faster than expected growth of non-fossil BGGs) lies with our sample all being strong gravitational lenses.  Since each of these systems are acting as strong lenses, they already must be concentrated systems thereby making each CASSOWARY group inherently more fossil-like than a typical non-lensing system.  This higher than average mass concentration has the potential to cause lensing systems to evolve differently than comparable non-lensing systems, and thus make direct comparisons to N-body models more difficult.

\begin{figure}
\hspace{-0.2truein}
{\includegraphics[scale=0.47]{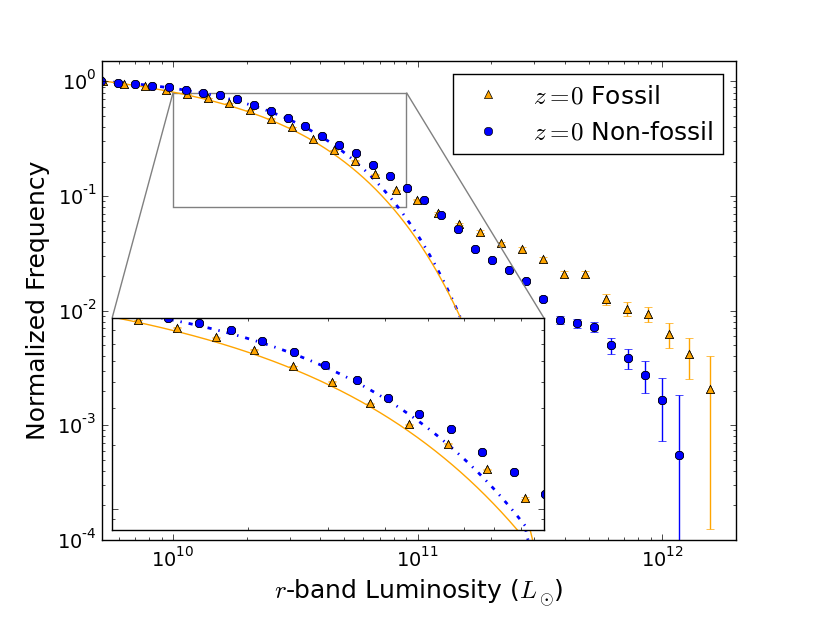}}
\caption{\footnotesize
{The projected $z=0$ galaxy luminosity functions of all eventual fossil systems (fossils + progenitors) found by numerically merging all member galaxies that could merge within each system's look back time to create a probable final fossil BGG.  The same was done for non-fossils for consistency in comparisons, although very little change was seen in its projected luminosity function.  The projected fossil deviation for $L_{gal}>10^{11}\;L_{\odot}$ galaxies is slightly more than what is observed at $z\sim0.4$, and a deficit in $10^{10}<L_{gal}<10^{11}\;L_{\odot}$ still present.}}
\label{fig:merged}
\end{figure}

We also begin to visually see the known $L^*$ galaxy ($10^{10}<L_{gal}<10^{11}\;L_{\odot}$) deficit in member galaxies in our projected $z=0$ fossils when compared to our projected $z=0$ non-fossils (Figure~\ref{fig:merged}). One can take the total light in this observed $10^{10}<L_{gal}<10^{11}\;L_{\odot}$ deficit of fossil systems and compare that to the total excess of light seen for $L_{gal}>10^{11}\;L_{\odot}$.  Since both populations (fossils + progenitors and non-fossils) have a comparable total stellar content (within a factor 1.6) we can see if the `missing' light at the faint end matches the `excess' seen a the bright end of the luminosity function for fossils.  We find that for both the rest frame (unmerged) and $z=0$ (merged) functions, these two deficits/excesses are within 7\% of each other supporting the idea that fossil BGGs aren't simply an over-luminous central galaxy.  Rather, they are more likely a product of the redistribution of the total group stellar content that has been focused into one galaxy, as this should preserve any differences in luminosity functions provided fossil and non-fossil BGGs grow in the same manner and at similar rates.

Fast forwarding each CASSOWARY system to $z=0$ yielded an unexpected finding where our projected $z=0$ non-fossil luminosity function matched our rest frame ($z\sim0.4$) `fossil-like' function to a surprising degree where a K-S test shows each coming from identical distributions.  This suggests that CASSOWARY fossils and progenitors are simply $\sim4\;$Gyr more evolved than non-fossils making age a defining factor in fossil systems.  These findings support the hypothesis that fossil systems are a phase of galaxy system evolution that all groups will eventually pass into or even through as new galaxies simultaneously fall into the group and existing galaxies are cannibalized by the BGG (von Benda-Beckmann et al.\ 2008; Cui et al.\ 2011).  However, this does not explain the higher than expected X-ray luminosities and temperatures seen in progenitors, indicating that they are not relaxed systems.  It is possible that this result is due to the strong lensing bias telling us that our lensing progenitors have a deeper potential well than non-lensing non-fossils of comparable richness.

\section{Individual Systems of Note}

\subsection{\textbf{CSWA 26}}

Our {\it Chandra} snapshots were designed to give the first definitive X-ray detections of eight fossil progenitors, however two systems were luminous enough for us to find additional information. This included rough radial temperature profiles which is a key component in identifying which formation track a specific fossil progenitor is following.  We see a clear asymmetry in CSWA 26's hot gas with a 50 kpc offset between the lensing center of mass and the X-ray centroid (Figure~\ref{fig:collage}).  We also observe a temperature increase at the group center going from $2.0^{+0.7}_{-0.4}$ keV in a  280-420 kpc annulus centered on the BGG, up to $6.8^{+1.7}_{-1.3}$ keV for the innermost circle of 280 kpc{\footnote[3]{All temperature errors are reported at the $90\%$ confidence level}}.  This factor of three increase in gas temperature is not consistent with a relaxed system but instead mirrors the previously studied fossil progenitor CSWA 2 (the Cheshire Cat) where it is believed a group merger was shock heating the gas at the center of the group (Irwin et al.\ 2015) making CSWA 26 a candidate for another group merger fossil progenitor but a factor of two to three times more massive.  This is supported optically by SDSS spectroscopic redshifts for the first ($z=0.336$) and third ($z=0.341$) rank member galaxies showing a $1030$ km$\;$s$^{-1}$ radial velocity difference.  For comparison, the radial velocity difference between the two `eyes' of the CSWA 2 is $1100$ km$\;$s$^{-1}$.  While SDSS photometric redshift (photoZ) data for all CSWA 26 member galaxies suggests a bimodality centered about the first and third rank galaxies, spectra are needed to confirm this as a second group merger fossil progenitor.

\subsection{\textbf{CSWA 28}}

Most unexpected of all was the $z=0.418$ progenitor CSWA 28 which exceeded X-ray luminosity predictions by an order of magnitude.  This optically unassuming $N_{200}=31$ fossil progenitor, by almost all accounts, seemed to be a relaxed system with little major merging left to complete before the fossil BGG was fully assembled.  Archival {\it HST} imaging showed smooth isophote contours in and around the BGG with the only oddity being a 62 kpc offset between the lensing center of mass and the BGG.  What was expected to be $\sim100$ net counts detection was instead over 1100 making CSWA 28 the most luminous fossil progenitor out of our sample.  CSWA 28 also shows the highest gas temperature of our sample at $7.1^{+1.8}_{-1.3}$ keV, which is expected from a massive cluster, not a poor fossil progenitor.  Like CSWA 26, we observe a radial temperature spike in the central regions of CSWA 28 from $3.0^{+0.7}_{-0.7}$ keV at 280 kpc to $9.2^{+5.3}_{-2.2}$ keV within 170 kpc meaning this could also be another group merger but in a post merger stage, as optical isophotes are smooth and no bright galaxies are nearby.  Interestingly, the BGG aligns well with the X-ray centroid; this means the lensing center of mass is being affected by something outside the system.  As with CSWA 26, there are currently no spectra of CSWA 28 making further investigation difficult.

\subsection{\textbf{CSWA 36}}

While CSWA 36 did not yield enough counts to identify the existence of any non-cool cores, the global temperature is higher than the $T_X-N_{200}$ relation by $4.8\sigma$, and we are able to see a significant elongation of the hot gas to the southwest of the BGG (Figure~\ref{fig:collage}).  There is also 60 kpc offset of the BGG to the X-ray centroid which could be, again, due to a group merger scenario where the BGG has been temporarily pulled away from the center of the dark matter potential; however to verify this, spectra are needed to search for the velocity distribution of all group members.  If CSWA 36 is confirmed to be another group merger fossil progenitor, that would mean $\sim45\%$ of the X-ray detected CASSOWARY progenitors (CSWA 2, 26, 28, and 30) show evidence of following the group merger fossil formation track suggesting this mechanism could contribute significantly to the total observed fossil fraction.

\section{Summary}
\subsection{\textbf{X-ray Conclusions}}
In this work, we find from {\it Chandra} ACIS-S snapshots of eight previously unobserved CASSOWARY fossil system progenitors systematic $L_X-T_X$, $L_X-N_{200}$, and $T_X-N_{200}$ relation offsets for progenitors making them brighter and hotter than what is expected for non-fossils of similar galaxy richness.  Progenitors are found to be between $2\sigma$ and $9\sigma$ removed from existing X-ray scaling relations, showing these deviations are statistically significant.  These offsets could be due to progenitors having deeper dark matter potential wells than non-fossils (leading to more retained hot gas at higher temperatures), progenitors undergoing group mergers which would shock heat gas and also introduce more X-ray gas into the system, or a bias caused by our targets being strong gravitational lenses.  We rule out the possibility that these temperature deviations are a result of $L^*$ galaxies being cannibalized by the BGG thus lowering $N_{200}$ and sliding our progenitors away from expectations by noting that an average of only two to three $L^*$ member will merge with the BGG by $z=0$.  While noticeable, cannibalization alone does not explain the $3\sigma$ average offset observed.

We observe that our progenitors are more consistent with fossil systems than non-fossils for relations involving hot gas temperature $T_X$ supporting the hypothesis that fossil-like systems are centrally concentrated before the fossil BGG has finished its assembly.  However, nearby fossils are not elevated on the $L_X-N_{200}$ relation while our progenitors are.  This suggests that the elevated $L_X$ we see for progenitors might be transitory and a result of their presently tumultuous environment possibly shock heating some of the gas.  It is also possible that AGN could inject a substantial amount of energy into the ICM thus temporarily increasing the group's X-ray luminosity.  However, we find it curious that out of nine X-ray detected fossil progenitors, only one (CSWA 2) shows evidence of an active AGN of $L_X>10^{41}\;$erg s$^{-1}$ in a member galaxy even though all possess congested central regions where major merging is either imminent or ongoing.

Asymmetries in X-ray emission for CSWA 26 and 36 and elevated core temperatures for CSWA 26 and 28 support the validity of a group merger mechanism for fossil formation and demonstrate that it may play a significant role in the total fossil fraction seen at low redshift today.  Finding tentative evidence for temperature spikes in CSWA 26 and 28 also aids in explaining the existence and unexpectedly high number of nearby fossil systems that do not possess cool cores, as X-ray cooling time scales are typically longer than galaxy merger time scales.  This means that these two progenitors should be classified as fossil systems before the hot gas at their cores has had a chance to cool significantly akin to the fossil progenitor CSWA 2.

\subsection{\textbf{{\it HST} Conclusions}}
Using new {\it HST} and archival observations of our eight strong lensing fossil progenitors, we are able to resolve the complex merging environments expected to exist in progenitors.  Two fossil systems identified via SDSS imaging (CSWA 4 and 11) were found to have unresolved $L^*$ galaxies within two magnitudes of their BGGs in mid cannibalization thereby shifting them from fossil to progenitor status.  Resolving more intermediate mass galaxies near progenitor BGGs such as these allows us to refine the galaxy luminosity functions for our fossil progenitors, the results of which support our previous findings that progenitors are the transition phase between non-fossils and fossils.  Interestingly, removing all BGGs from the data cause our fossils, progenitors, and non-fossils to become consistent with one another, however the expected differences from non-fossils appeared when the fossil and progenitor samples were combined.  We note that the reclassifying of two fossils to progenitors caused fossil luminosity function uncertainties to grow substantially possibly washing out subtle differences such as these.  We also note that our data were limited to $M_r<-19.5$, as this is the SDSS completeness threshold for our most distant system.  As much debate surrounds the faint end behavior of the fossil luminosity function ($M_r>-18.0$), it is possible that differences between our fossils and non-fossils exists outside the BGGs but in a place we are unable to probe with this data set.

By combining galaxies from the fossil and progenitor categories to form a `fossil-like' class of groups, we find the expected deficit $10^{10}<L_{gal}<10^{11}\;L_{\odot}$ members whose combined luminosity matches within 7\% the excess luminosity seen in fossils for $L_{gal}>10^{11}\;L_{\odot}$.  This is preserved even after fast forwarding each system to a $z=0$ reference frame by numerically merging eligible member galaxies according to the system's look back time and the merger time scale from Kitzbichler \& White (2008).  We also notice that the `fossil-like' and non-fossil luminosity functions evolve together when placed in a $z=0$ frame, implying that their BGGs are growing at comparable rates which appears to contradict findings in the Millennium simulation by Gozaliasl et al.\ (2014).  At the same time, comparable BGG growth rates for our fossils and non-fossils requires our fossil BGGs to be assembled earlier than the non-fossils', otherwise the known fossil luminosity function deviation at the bright end would never form.  This early fossil BGG assembly time supports other findings from Gozaliasl et al.\ (2014) who saw most of the fossil BGG mass assembly occur before $z=0.5$.

\acknowledgments
Acknowledgements:  We thank William Keel, Jeremy Bailin, Spring Johnson, and Eddie Johnson for their insightful comments and suggestions regarding this work.  This work was supported by {\it Chandra} grant GO6-17106X and {\it HST} grant HST-GO-14362.003-A.

\begin{footnotesize}

\end{footnotesize}

\end{document}